\RequirePackage[2020-02-02]{latexrelease}
\documentclass[preprint,aps,floatfix]{revtex4}
\DeclareUnicodeCharacter{00A0}{ }
\usepackage{graphicx}
\usepackage{textgreek}
\usepackage{verbatim} 
\usepackage{amssymb}
\usepackage{url}
\usepackage{amsmath,amssymb}
\usepackage{mathtools} 
\usepackage[symbol]{footmisc}
\usepackage{subcaption}
\usepackage{capt-of}
\usepackage{bigints}
\usepackage{romannum}
\def\nn{\nonumber}
\AtBeginDocument{\pagenumbering{arabic}}
\begin{document}

\title{Interacting tachyonic scalar field \Romannum{2}}

  \author{V K Ojha${}^1$\footnote{vko@phy.svnit.ac.in}, Adithya A Rao${}^1$\footnote{adithyarao3132001@gmail.com}, and S D Pathak${}^2$	\footnote{prince.pathak19@gmail.com}}
  
  \affiliation{${}^1$ Department of Physics, Sardar Vallabhbhai National Institute of Technology, Surat, 395 007, India.\\ ${}^2$Department of Physics, Lovely Professional University, Phagwara, Punjab, 144 411, India.}

   \begin{abstract}
    The existence of dark energy is essential to explain the cosmic accelerated expansion.  We consider a homogenous interacting tachyonic scalar field as a possible candidate for the dynamical dark energy. The interaction between the tachyonic field and matter can be gauged to be linear in the energy density of matter (or the tachyonic field) and Hubble's parameter. We estimate the rate of expansion, the age of the universe, the evolution of energy density of matter and tachyonic field, and the coupling strength of the interaction for a spatially flat ($k=0$) universe. We observed that the upper limit of coupling strength is 1, and it is the same whether the interaction term depends on the energy density of matter or the energy density of tachyonic scalar field.
  
   \end{abstract}
   
  \maketitle


  \section{Introduction}
\noindent
 Type Ia supernovae \cite{Riess_1998,Perlmutter_1999} observation confirms the universe's accelerated expansion. This observation demands the existence of a medium with negative pressure, usually called dark energy. Different scalar fields are observed to satisfy this negative pressure property and have been used extensively to study dark energy \cite{copeland2006dynamics,padmanabhan2002can,verma2014cosmic,verma2014bicep2}. Some popular choices of scalar fields are phantom \cite{cruz2022phantom,narawade2022phantom,tripathy2020phantom,johri2004phantom,dkabrowski2006quantum,dabrowski2003phantom}, quintessence \cite{adil2023quintessential,yang2019constraints,wetterich2022quantum,piedipalumbo2020noether}, and tachyon \cite{felegary2020evolution,singh2019low,koussour2022interacting,sadeghi2014phenomenological,gorini2004tachyons,kundu2021interacting}. All these scalar fields exhibit negative pressure, making each a possible candidate for dark energy. In fact in \cite{Kumar:2023qum}, it has been shown that these three scalar fields are indistinguishable under the slow roll approximation. In this article, we consider the tachyonic scalar field as a candidate for dark energy, making it the source of accelerated expansion of the universe.

The tachyonic scalar field was first defined in string theory \cite{Sen_2002,Sen1_2002,Sen2_2002} and later used as a possible candidate for dark energy \cite{padmanabhan2002can,sadeghi2014phenomenological}. The equation of state for a tachyonic scalar field, $p = - \rho$, satisfies the negative pressure condition for a universe component necessary to be termed as dark energy. Without interactions, the tachyonic scalar field exhibits behavior
similar to the cosmological constant. But a static model of the universe, where the tachyonic field does not interact with other universe components, suffers from two major problems, the coincidence problem, and the Cosmological Constant problem. An interacting model of dark energy resolves this problem \cite{Chimento_2010,Chimento_2008,Bertolami_2012,Wang_2007,lu2012investigate,Farajollahi_2012,Zimdahl_2012,Yang_2018,Cao_2011,Di_Valentino_2020,Verma_2012,Verma_2013,V_liviita_2010,Pan_2018,Amendola_2018,B_gu__2019,Pan_2019,Papagiannopoulos_2020,Savastano_2019,von_Marttens_2019,yang2019reconstructing,Asghari_2019}. In such a model, the dark energy field is considered dynamic, with energy exchange between the matter and the universe's dark energy content. A major challenge in considering the interacting model is fixing the interactions' functional forms. Different forms of interactions have been proposed by several authors based on dimensional and phenomenological arguments over the years \cite{Chimento_2010,Chimento_2008,Bertolami_2012,Wang_2007,lu2012investigate,Farajollahi_2012,Zimdahl_2012,Yang_2018,Di_Valentino_2020,shahalam2017dynamics,shahalam2015dynamics}.

We consider an interacting model of dark energy, in which the dark energy is modeled by a tachyonic scalar field that interacts with the universe's matter content. Specifically, we consider that the interaction between the tachyonic field and matter depends linearly on the energy density of matter. Using this interacting model, we investigate the behavior of the energy density, scale factor, and age of the universe and the possible values of the coupling strength. Previously we had considered the interaction to be linearly dependent on the tachyonic field and found that the coupling strength of interaction can not exceed the value 1 \cite{kundu2021interacting}.

The article is organized as follows. We start with some theoretical background in the section
\ref{background}, and then briefly discuss the interacting tachyonic scalar field model in section \ref{ITSF}. Section
\ref {sec: energydensity} and \ref{sec:omega} consist of derivation and analysis of the evolution of energy density for matter and
field. The evolution of scale factor and the age of the universe are discussed in section \ref{sec: scale}, and \ref{sec: aou} respectively. A comparison of coupling strength when interaction depends linearly on matter
density and tachyonic field is presented in section \ref{sec: aplha vs beta}. Finally, we conclude in section \ref{sec: Con}.

    \section{Theoretical Background}\label{background}
\noindent
    The FLRW metric in natural unit is given by 
    \begin{equation}
        g^{\mu\nu} = \text{diag}\left( -1,~ \frac{a^2}{1 - \displaystyle\frac{\kappa r^2}{R_0^2}},~a^2r^2,~ a^2r^2\sin^2\theta \right)
    \end{equation}
    where $a$ is the time-dependent scale factor, and $\kappa$ is the global curvature of the universe. This metric follows from the cosmological principle, which implies that for our universe, the stress-energy tensor takes on a relatively simple form 
    \begin{equation}
        T^{\mu\nu} = \text{diag}( \rho(t)^2,~ p(t),~ p(t),~ p(t) )
    \end{equation}
    For a universe with the FLRW metric and the stress energy tensor given above, the Friedmann equation follows from the Einstein field equations and is given as 
    \begin{equation}
        \left( \frac{\dot{a}(t)}{a(t)} \right)^2 = \frac{8\pi G}{3c^2}\rho(t) - \frac{\kappa c^2}{R_0^2}a(t)^2
    \end{equation}
    For a flat universe, $\kappa = 0$, and the Friedmann equation takes on a simpler form 
    \begin{equation}
        \left( \frac{\dot{a}(t)}{a(t)} \right)^2 = \frac{8\pi G}{3c^2}\rho(t)
    \end{equation}
    The principle of conservation of energy, when applied to the energy component of the universe gives the fluid equation
    \begin{equation}
        \dot{\rho}(t)  + 3\frac{\dot{a}(t)}{a(t)}\left(\rho(t) + p(t)\right) = 0
    \end{equation}
    The last two equations, along with the equation of state, $p(t)=\omega \rho(t)$, specify the dynamics of the universe. 
%
%
    \section{Interacting Tachyonic Scalar Field}\label{ITSF}
\noindent 
    We consider the universe with two dominant components: matter and dark energy, with dark energy modeled by a spatially homogenous tachyonic scalar field (TSF). The Lagrangian density for TSF is \cite{Sen_2002,Sen1_2002}
    \begin{equation}
        \mathcal{L} = -V(\phi) \sqrt{1 - \partial^\mu \partial_\mu \phi}.
    \end{equation}
    For such a TSF, the stress-energy tensor is
    \begin{equation}
        T^{\mu\nu} = \frac{\partial \mathcal{L}}{\partial (\partial_{\mu}\phi) }\partial^{\nu}\phi - g^{\mu\nu}\mathcal{L},
    \end{equation}
    and the energy and pressure density follows directly from the stress-energy tensor as
    \begin{equation}
        \rho = \frac{V(\phi)}{\sqrt{1 - \partial^\mu \partial_\mu \phi}},
    \end{equation}
    and
    \begin{equation}
        p = -V(\phi)\sqrt{1- \partial^\mu \partial_\mu \phi}.
    \end{equation}
    Since the TSF is spatially homogeneous, the spatial derivatives vanish, while the time derivative survives. Thus
    \begin{equation}
        p = - (1 - \dot{\phi}^2)\rho
    \end{equation}
    giving $\omega_\phi = -(1-\dot{\phi}^2)$.\\
    \renewcommand{\thefootnote}{\fnsymbol{footnote}}
    \setcounter{footnote}{0}

    We consider the interaction of the TSF with matter via the transfer of energy. The two components exchange energy and hence violate individual energy conservation. But overall, the total energy of the universe is conserved. Thus for such a model, the continuity equation for the energy density of TSF ($\rho_\phi$) and matter ($\rho_m$) gets modified as 
    \begin{equation}
        \dot{\rho_\phi} + 3\frac{\dot{a}}{a}(1 + \omega_\phi)\rho_\phi = -Q,
    \end{equation}
    \begin{equation}
        \dot{\rho_m} + 3\frac{\dot{a}}{a}(1 + \omega_m)\rho_m =Q,
    \end{equation}
    where $\omega_\phi = p_\phi/\rho_\phi$, and $\omega_m = p_m/\rho_m$.\\
    Based on the phenomenological argument, the functional form of the interaction term can be guessed to be linear in the energy density of the matter or field \cite{Pav_n_2008,DeArcia:2022nps}. Previously we investigated the case when $Q$ depends linearly on the energy density of TSF ($\rho_\phi$) \cite{kundu2021interacting}. In this work, we investigate the case when $Q$ depends linearly on the energy density of matter ($\rho_m$). In particular, we choose $Q \propto \rho_m$, with the proportionality constant being $3\displaystyle\beta\frac{\dot{a}}{a}$, where $\beta$ is a dimensionless coupling constant specifying the strength of interaction. Thus the interaction term takes on the form
    \begin{equation}
        Q = 3\beta \frac{\dot{a}}{a} \rho_m
    \end{equation}


\subsection{Evolution of Energy Densities}
\label{sec: energydensity}
\noindent    
    With the above-said interaction term, the continuity equations read
    \begin{equation}\label{eq:ce1}
        \dot{\rho_\phi} + 3\frac{\dot{a}}{a}(1 + \omega_\phi)\rho_\phi = -3\beta \frac{\dot{a}}{a}\rho_m,
    \end{equation}
    and
    \begin{equation}\label{eq:ce2}
        \dot{\rho_m} + 3\frac{\dot{a}}{a}(1 + \omega_m)\rho_m = 3\beta\frac{\dot{a}}{a}\rho_m.
    \end{equation}
    These can be solved to get the equation governing the evolution of $\rho_\phi$ and $\rho_m$ with scale factor $a$.\\
\noindent
    $\bullet$ \textbf{Solving for $\rho_m$}
    \begin{equation}
        \dot{\rho_m} + 3\frac{\dot{a}}{a}(1+\omega_m) = 3\beta \displaystyle\frac{\dot{a}}{a}\rho_m, \nn
    \end{equation}
    \begin{equation}
        \implies \dot{\rho_m} + 3\frac{\dot{a}}{a}(1+\omega_m - \beta)\rho_m = 0
    \end{equation}
    This differential equation has a simple solution
    \begin{equation} \label{eq:DensityEvolutionMatter}
        \frac{\rho_m}{\rho_m^0} = \left(\frac{a}{a^0}\right)^{-3(1+\omega_m - \beta)}=  \left(\frac{a}{a^0}\right)^{-\gamma}
    \end{equation}
    where $\rho_m^0$ is the energy density of matter at the present time, and $a^0$ is the scale factor at present time.\\
    
\begin{figure}[h]
        \centering
        $\displaystyle\frac{\rho}{\rho^0}$ vs scale factor $a$\vspace{10pt}\\
        \begin{subfigure}{0.46\textwidth}
            \centering
            \includegraphics[width=\textwidth]{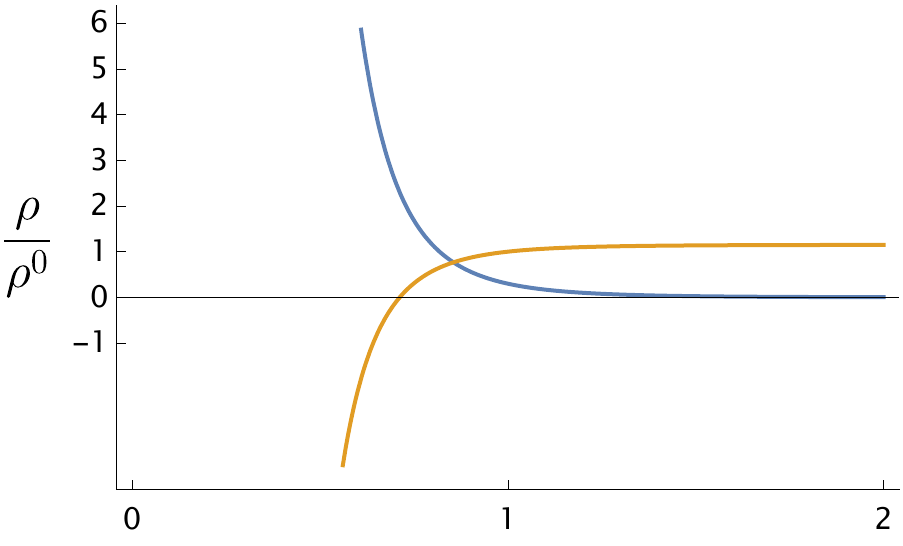}
            \caption{$\beta = -1$}
        \end{subfigure}
        \begin{subfigure}{0.5\textwidth}
            \centering
            \includegraphics[width=\textwidth]{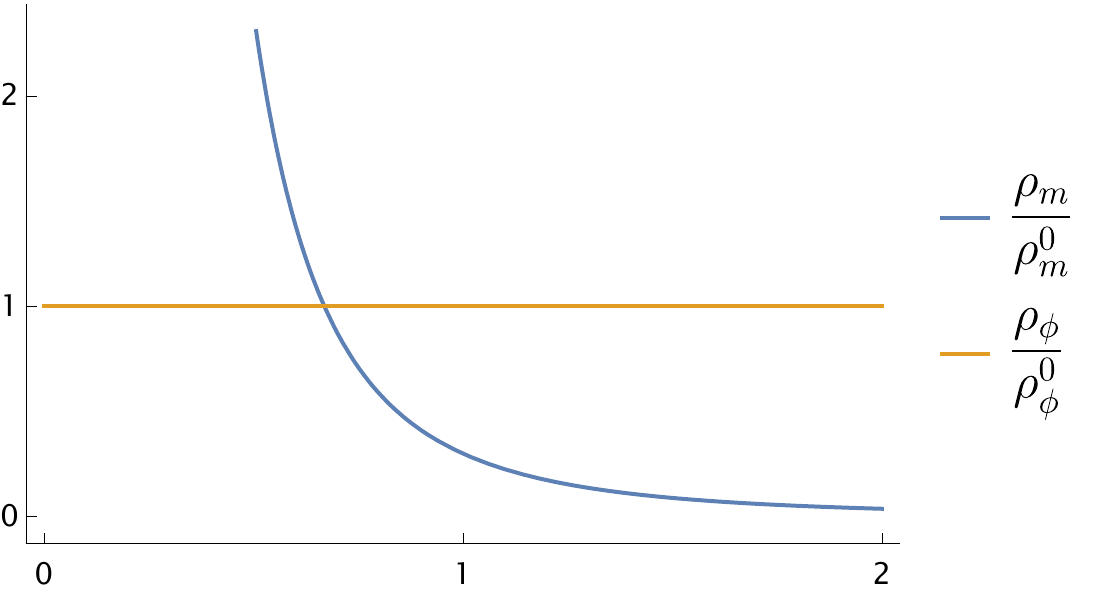}
            \caption{$\beta = 0$}
        \end{subfigure}
        \begin{subfigure}{0.46\textwidth}
            \centering
            \includegraphics[width=\textwidth]{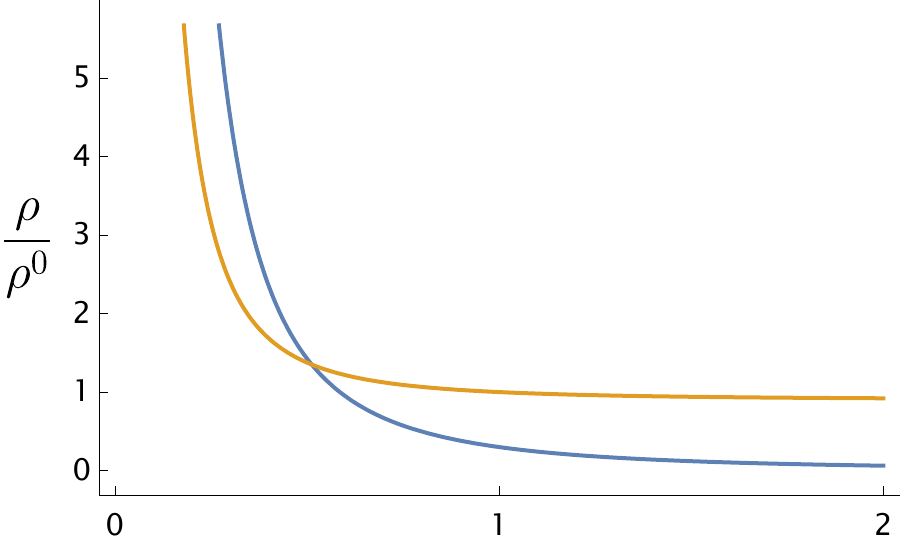}
            \caption{$\beta = 0.25$}
        \end{subfigure}
        \begin{subfigure}{0.5\textwidth}
            \centering
            \includegraphics[width=\textwidth]{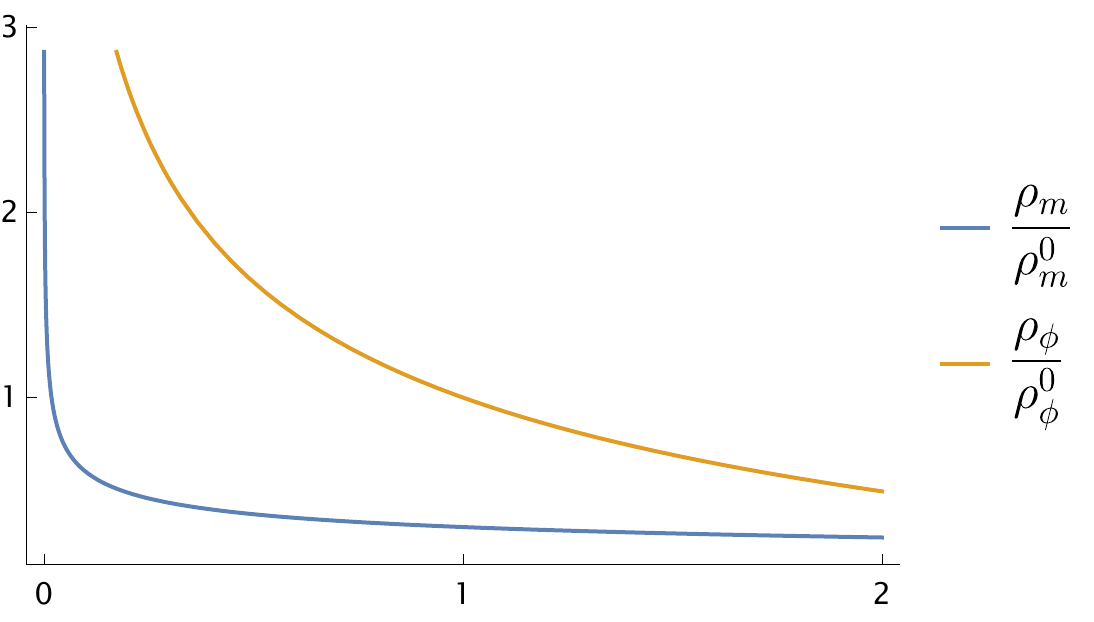}
            \caption{$\beta = 0.9$}
        \end{subfigure}
        \caption{Plot of the energy densities of matter and TSF against the scale factor for different values of coupling constant $\beta$.}\label{fig: densities}
    \end{figure}

    \noindent
    $\bullet$ \textbf{Solving for $\rho_\phi$}
    \begin{equation}
        \dot{\rho_\phi} + 3\frac{\dot{a}}{a}(1+\omega_\phi)\rho_\phi = -3\beta \displaystyle\frac{\dot{a}}{a}\rho_m^0 \left( \frac{a}{a^0} \right)^{-\gamma}
    \end{equation}
    Changing variables, $a = xa^0, ~\rho_\phi = R\rho_\phi^0 \implies \dot{a} = \dot{x}a^0,~ \dot{\rho_\phi} = \dot{R}\rho_\phi^0$

    \begin{equation}
        \dot{R} + 3\frac{\dot{x}}{x} (1+\omega_\phi)R = -3\beta \dot{x}\frac{\rho_m^0}{\rho_\phi^0}x^{-\gamma-1}
    \end{equation}
    The equation can be rewritten as
    \begin{equation}
        \frac{dR}{dx} + \frac{3(1+\omega_\phi)}{x}R = -3\beta \frac{\rho_m^0}{\rho_\phi^0} x^{-\gamma-1}
    \end{equation}
    Solving, we get
    \begin{equation}
            R = -x^{-3(1+\omega_\phi)}\left( 3\beta \frac{\rho_m^0}{\rho_\phi^0}\frac{x^{3(1+\omega_\phi)-\gamma}}{3(1+\omega_\phi)-\gamma} - 1 \right) - 3\beta \frac{\rho_m^0}{\rho_\phi^0}\frac{-x^{-3(1+\omega_\phi)}}{3(1+\omega_\phi)-\gamma}
    \end{equation}
    For the TSF to mimic the cosmological constant, $\omega_\phi = -1$ (implying that the TSF is a constant over time also, i.e. $\dot{\phi} = 0$), thus the above equation becomes
    \begin{equation}\label{eq:DensityEvolutionField}
        \frac{\rho_\phi}{\rho_\phi^0} = 3\beta \frac{\rho_m^0}{\rho_\phi^0}\frac{1}{\gamma}\left( \left( \frac{a}{a^0} \right)^{-\gamma} - 1\right) + 1
    \end{equation}
    The variation of the energy densities of matter and dark energy with the scale factor for different values of $\beta$ is plotted in Fig.(\ref{fig: densities}).

    It is worth noting that $\beta$ determines the coupling strength between matter and the scalar field $\phi$. For $\beta > 0$, dark energy decreases as matter density increases. On the other hand, for $\beta < 0$, dark energy increases as matter density increases. When $\beta = 0$, the interaction vanishes, and the two sectors evolve independently, i.e., the cosmological constant scenario.



\subsection{Evolution of $\Omega$ with $\text{ln}(a)$}
\label{sec:omega}
\begin{figure}[h]
            \centering
            $\Omega~\text{vs ln}(a)$ for different values of coupling constant $\beta$ \vspace{10pt}\\
            \begin{subfigure}{0.42\textwidth}
                \centering
                \includegraphics[width=\textwidth]{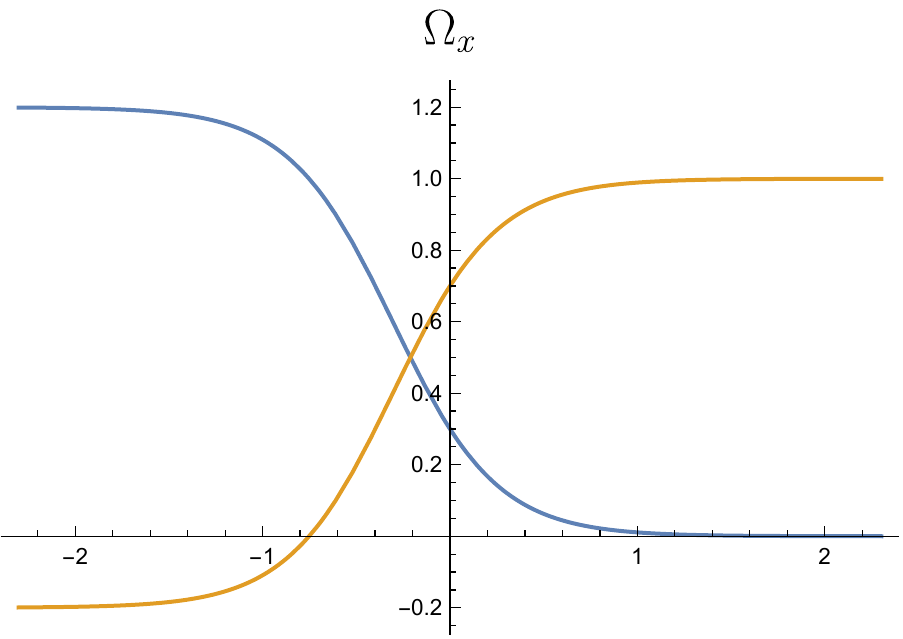}
                \caption{$\beta = -0.2$}
            \end{subfigure}
            \begin{subfigure}{0.57\textwidth}
                \centering
                \includegraphics[width=\textwidth]{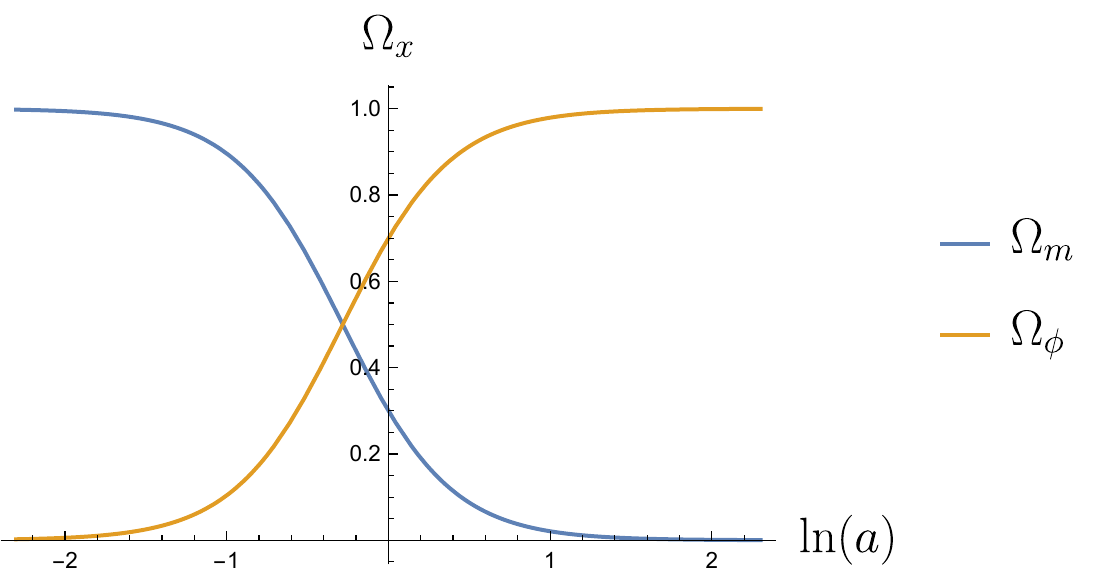}
                \caption{$\beta = 0$}
            \end{subfigure}
            \begin{subfigure}{0.42\textwidth}
                \centering
                \includegraphics[width=\textwidth]{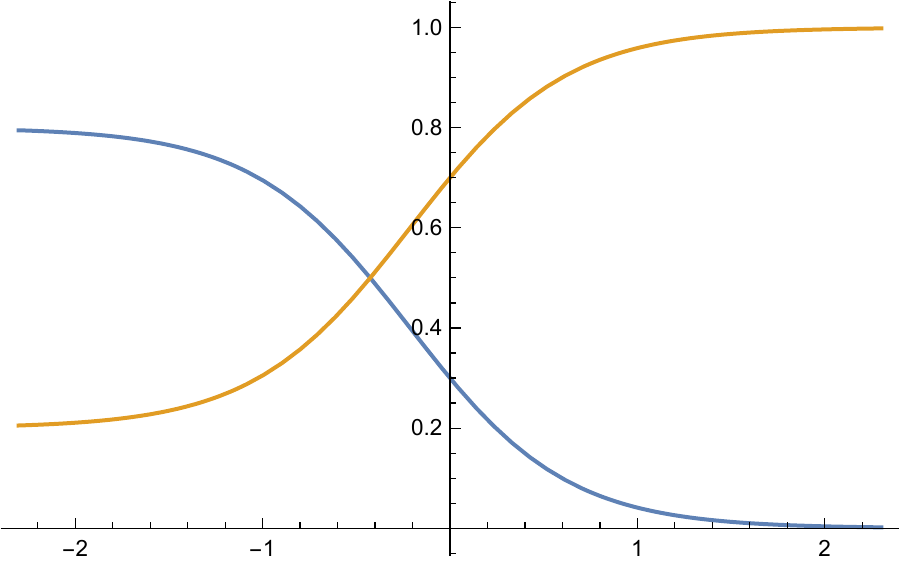}
                \caption{$\beta = 0.2$}
            \end{subfigure}
            \begin{subfigure}{0.57\textwidth}
                \centering
                \includegraphics[width=\textwidth]{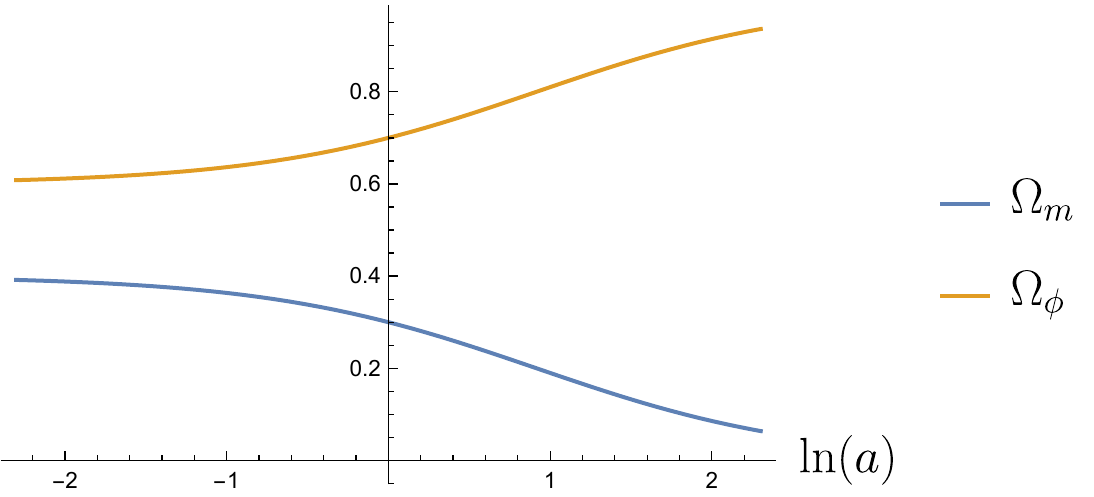}
                \caption{$\beta = 0.6$}
            \end{subfigure}
            \caption{Plot of the evolution of density parameter $\Omega$ for matter and TSF as the function of ln$(a)$ for different values of coupling constant $\beta$.}
            \label{fig: omega}
        \end{figure}

\noindent
        The equations of cosmology can be written in terms of another variable - density parameter, defined as the ratio of the energy density and the critical density $\rho_c$,
        \begin{equation}
            \Omega = \frac{\rho}{\rho_c} = \frac{8\pi G a^2}{3\dot{a}^2}\rho = \frac{8\pi G}{3H^2}\rho,~\text{with}~H = \frac{\dot{a}}{a}.
        \end{equation}
        The numerical value of $\Omega$ indicates the nature of the universe locally.
        For $\Omega = 1$, the universe is spatially flat; for $\Omega<1$, it is negatively curved; and for $\Omega >1$, it is positively curved.\\
        Since, for any function $f$
        \begin{equation}
            \frac{df}{d\text{ln}(a)} = \frac{df}{da}/\frac{d\text{ln}(a)}{da} = a.(\frac{df}{dt}/\frac{da}{dt}),\nn
        \end{equation}
        therefore, for $H = \displaystyle\frac{\dot{a}}{a}$, we have
        \begin{equation}
            \frac{df}{d\text{ln}(a)} = \frac{1}{H} \frac{df}{dt}.
        \end{equation}
        Defining $\Omega_m = \displaystyle\frac{\rho_m}{\rho_c}$, $\Omega_\phi= \displaystyle\frac{\rho_\phi}{\rho_c}$, we can write
        \begin{equation}\label{eq:DensityParameterDE1}
            \frac{d\Omega_m}{d\text{ln}(a)} = \frac{8\pi G}{3}\frac{1}{H}\left( \frac{\dot{\rho_m}}{H^2} -2\frac{\rho_m}{H^3}\dot{H}\right)
        \end{equation}
        and
        \begin{equation}\label{eq:DensityParameterDE2}
            \frac{d\Omega_\phi}{d\text{ln}(a)} = \frac{8\pi G}{3}\frac{1}{H}\left( \frac{\dot{\rho_\phi}}{H^2} -2\frac{\rho_\phi}{H^3}\dot{H}\right)
        \end{equation}
        From the continuity equations (Eq.(\ref{eq:ce1},\ref{eq:ce2})), we get
        \begin{equation}\label{eq:DensityRate}
            \dot\rho_\phi = -3\beta H \rho_m~~\&~~ \dot\rho_m = -3H\rho_m + 3\beta H \rho_m 
        \end{equation} 
        and from the Friedmann equation, we get
        \begin{equation}\label{eq:HubbleRate}
            2H\dot{H} = \frac{8\pi G}{3}(\dot\rho_m + \dot\rho_\phi) = \frac{8\pi G}{3}(-3\rho_m)\nn
        \end{equation}
        \begin{equation}
            \implies \dot{H}= \frac{-8\pi G \rho_m}{2}
        \end{equation}
        Using Eqs.(\ref{eq:DensityRate},\ref{eq:HubbleRate}), and the fact that for a spatially flat universe such as ours, $\Omega_m + \Omega_\rho = 1$, the differential Eqs.(\ref{eq:DensityParameterDE1},\ref{eq:DensityParameterDE2}) become
        \begin{equation}
            \frac{d\Omega_m}{d\text{ln}(a)} = 3\beta \Omega_m + 3 \Omega_m^2 -3\Omega_m,
        \end{equation}
        and 
        \begin{equation}
            \frac{d\Omega_\phi}{\text{ln}(a)} =3 (1-\Omega_\phi)\Omega_\phi -3\beta (1-\Omega_\phi).
        \end{equation}
        These equations are solved, using conditions $\Omega_m |_{a = 1} = 0.3$ and $\Omega_\phi |_{a = 1} = 0.7$ for different values of $\beta$ and the evolution of $\Omega_m$ and $\Omega_\phi$ with ln$(a)$ is shown in Fig.(\ref{fig: omega}).\\

        For $\beta=0$ we recover the result for our universe, the matter density is dominant initially, and in the future, the dark energy density will be dominant. Also we can see that we are at the epoch of equality of $\Omega_m$ and $\Omega_\phi$. For negative $\beta$, $\Omega_\phi$ goes to negative values as expected from the behavior of $\rho_\phi$. For universes with positive $\beta$, the initial difference between $\Omega_m$ and $\Omega_\phi$ is reduced as $\beta$ is increased, and for $\beta \gtrapprox 0.5$, dark energy becomes the dominant constituent at all times.
        
    \section{Evolution of scale factor}
    \label{sec: scale}
\begin{figure}[h]
        \centering
        $a~\text{vs}~ t$ for different values of coupling constant $\beta$ \vspace{10pt}\\
        \begin{subfigure}{0.5\textwidth}
            \centering
            \includegraphics[width=\textwidth]{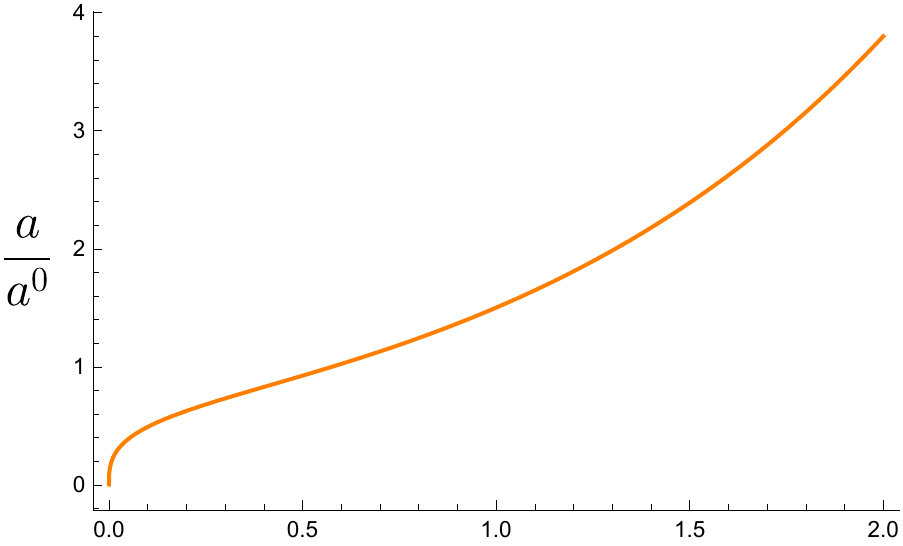}
            \caption{$\beta = -1$}
        \end{subfigure}
        \begin{subfigure}{0.46\textwidth}
            \centering
            \includegraphics[width=\textwidth]{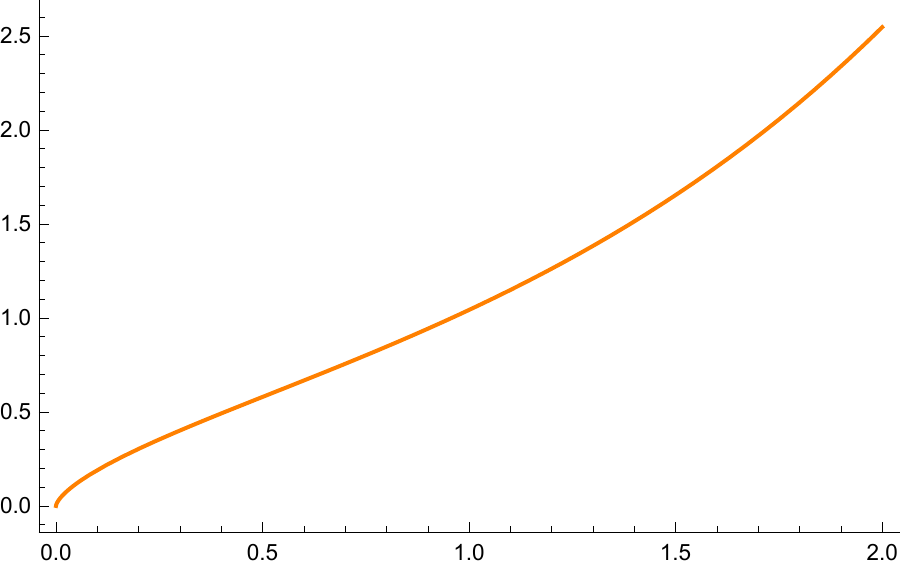}
            \caption{$\beta = 0$}
        \end{subfigure}
        \begin{subfigure}{0.5\textwidth}
            \centering
            \includegraphics[width=\textwidth]{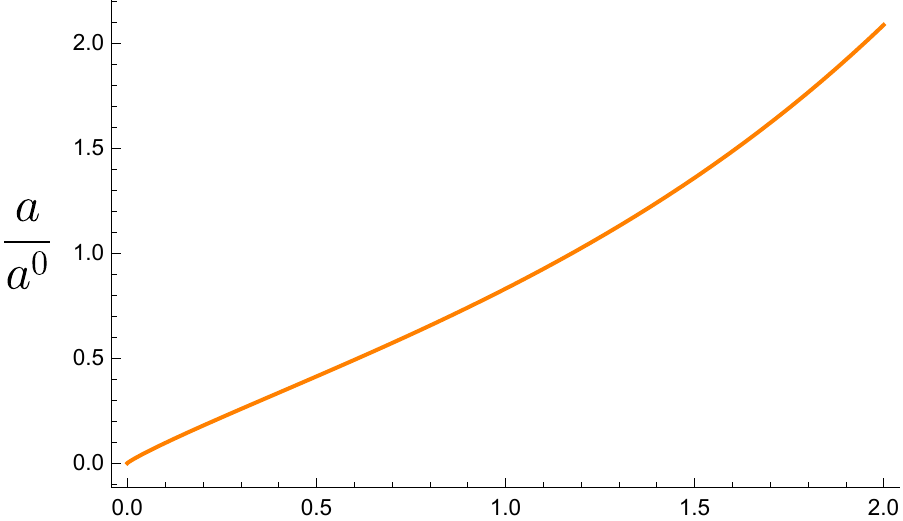}
            \caption{$\beta = 0.25$}
        \end{subfigure}
        \begin{subfigure}{0.46\textwidth}
            \centering
            \includegraphics[width=\textwidth]{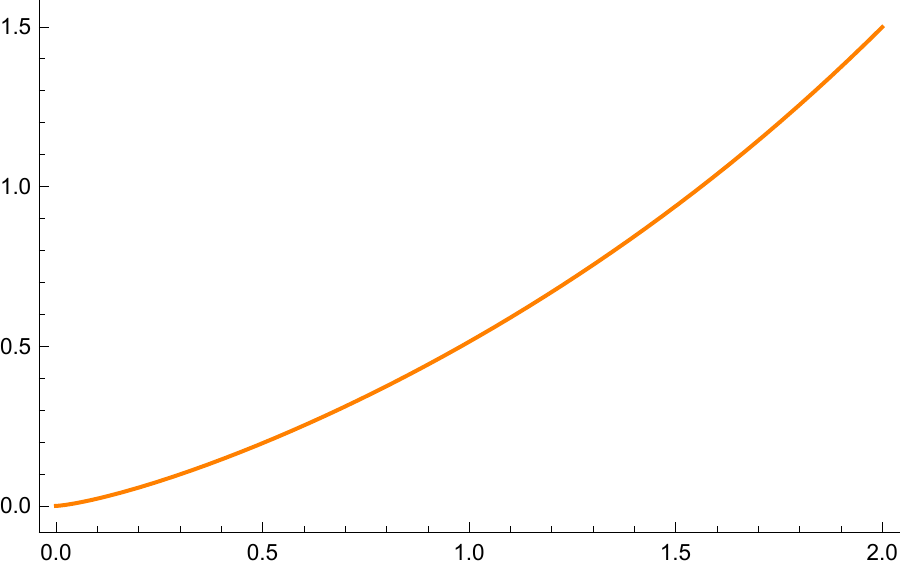}
            \caption{$\beta = 0.5$}
        \end{subfigure}
        \begin{subfigure}{0.48\textwidth}
            \centering
            \includegraphics[width=\textwidth]{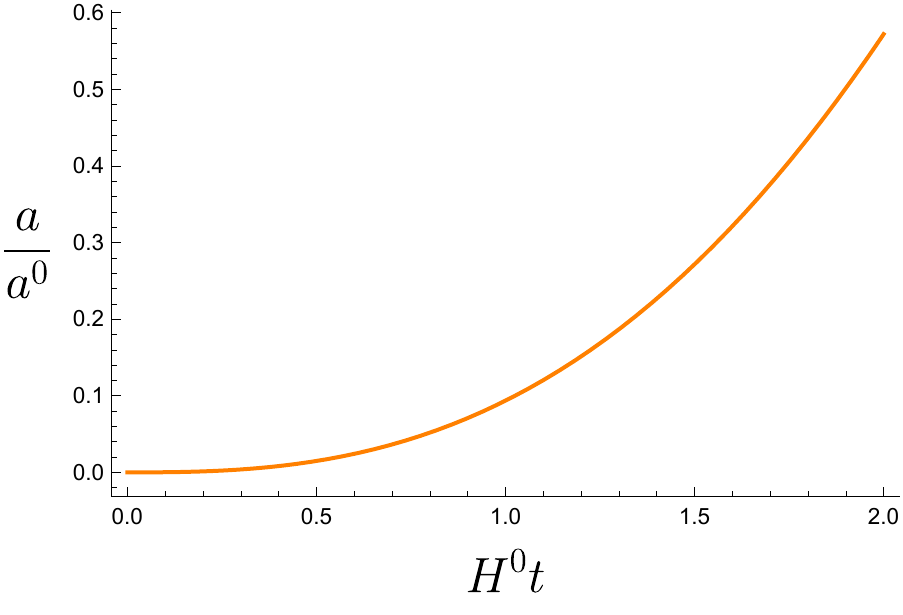}
            \caption{$\beta = 0.75$}
        \end{subfigure}
        \begin{subfigure}{0.48\textwidth}
            \centering
            \includegraphics[width=\textwidth]{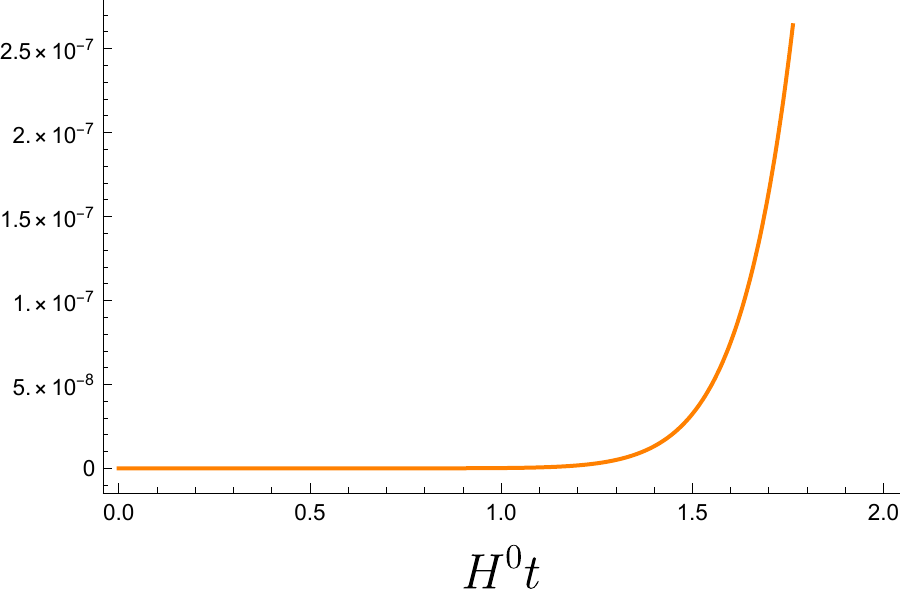}
            \caption{$\beta = 0.95$}
        \end{subfigure}
        \caption{Plot of the evolution of scale factor $a$ of the universe for different values of coupling constant $\beta$.}
        \label{fig: scale}
    \end{figure}

\noindent
    Another important  factor to consider is how the scale factor varies with time in the universe. This not only quantifies the expansion of the universe but also allows for the conversion between functions of $a$ and $t$. The Friedmann equation gives a relation between $a$ and $\rho$ and this can be used to obtain the form of scale factor as a function of time.
    The Friedmann equation can be written as
    \begin{equation}
        \frac{\dot{a}}{a} = \sqrt{\frac{8\pi G}{3c^2}} ~\sqrt{\rho_\phi + \rho_m} .
    \end{equation}
    Using the equation for the evolution of energy densities (Eqs.(\ref{eq:DensityEvolutionMatter},\ref{eq:DensityEvolutionField})), and $a = xa^0$, we get
    \begin{equation}
        \frac{\dot{x}}{x} = \sqrt{\frac{8\pi G}{3c^2}} \sqrt{\rho_m^0 x^{-\gamma} + \rho_\phi^0 \left(3\beta \frac{\rho_m^0}{\rho_\phi^0}\frac{1}{\gamma}\left( x^{-\gamma} - 1\right) + 1 \right) },\nn
    \end{equation}
    which on further simplification becomes
    \begin{equation}
        \frac{\dot{x}}{x} = \sqrt{\frac{8\pi G\rho_m^0}{3c^2}} \sqrt{ x^{-\gamma} + 3\beta \frac{1}{\gamma}\left( x^{-\gamma} - 1\right) + \frac{\rho_\phi^0}{\rho_m^0} }.
    \end{equation}
    From this, we can write
    \begin{equation}
        t = \sqrt{\frac{3c^2}{8\pi G\rho_m^0}} \bigint \frac{dx}{  x \sqrt{\displaystyle \frac{1}{x^{\gamma}} \left(1 + \frac{3\beta}{\gamma} \right) +\left(\frac{\rho_\phi^0}{\rho_m^0} -\frac{3\beta }{\gamma}\right) }  ~},
    \end{equation}
    and using $\omega_m = 0$, $\rho_c^0=8.7\times 10^{-27} ~\mathrm{kg/m^3}$, $\rho_m^0=\Omega_m^0 \rho_c^0$, $\rho_\phi^0=\Omega_\phi^0 \rho_c^0$  and integrating over $x$ gives 

    \begin{equation}
        H^0 t = -\frac{6.28701 \sinh ^{-1}\left(\displaystyle\frac{1}{3} \sqrt{21-30 \beta} \sqrt{x^{3-3\beta}}\right)}{\sqrt{21-30 \beta } \sqrt{3-3\beta }}\nn,
    \end{equation}
    implying that the scale factor $x$ evolves as

    \begin{equation}
        x = \frac{a}{a^0} = \left(3\frac{\sinh \left(0.159 \sqrt{21-30 \beta} \sqrt{3-3\beta} ~tH^0\right)}{\sqrt{21-30 \beta}}\right)^{\displaystyle\frac{2}{3-3\beta}}.
    \end{equation}
    The expansion of the universe, i.e., the evolution of the scale factor with time for different values of coupling constant $\beta$ is plotted in FIG. \ref{fig: scale}.
    For our universe, with $\beta = 0$, the plot shows that the universe's expansion is decelerating in the initial stage and then accelerating at the later stage. For negative $\beta$, the initial expansion rate of the universe is magnified, making the deceleration curve more prominent. In contrast, for positive $\beta$, the initial expansion rate is very small, and the universe's deceleration does not occur.
    

    \section{Age of the Universe}
    \label{sec: aou}
\begin{figure}[h]
        \begin{minipage}[b]{0.5\textwidth}
                \includegraphics[width=\textwidth]{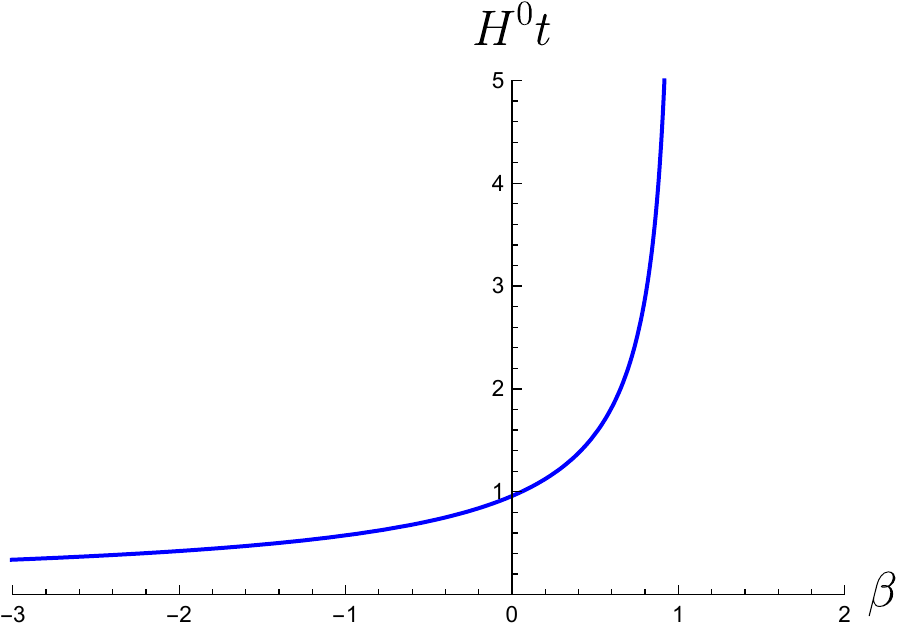}
                \captionof{figure}{The Age of the Universe (AOU) as the function of coupling constant.}
                \label{fig:AOUvsbeta}
        \end{minipage}
        \hspace{0.05\textwidth}
        \begin{minipage}[b]{0.4\textwidth}
            \begin{center}
                \renewcommand*{\arraystretch}{1.15}
                \begin{tabular}{c|c}
                    \begin{tabular}{cc}
                        $\beta$ & AOU \\\hline
                        -1. & 0.575731 \\
                        -0.9 & 0.598269 \\
                        -0.8 & 0.622909 \\
                        -0.7 & 0.649977 \\
                        -0.6 & 0.679872 \\
                        -0.5 & 0.713089 \\
                        -0.4 & 0.75025 \\
                        -0.3 & 0.792149 \\
                        -0.2 & 0.839813 \\
                        -0.1 & 0.894606 \\
                    \end{tabular} & 
                    \begin{tabular}{cc}
                        $\beta$ & AOU \\\hline
                        0. & 0.958375 \\
                        0.1 & 1.03369 \\
                        0.2 & 1.12427 \\
                        0.3 & 1.23568 \\
                        0.4 & 1.37672 \\
                        0.5 & 1.56232 \\
                        0.6 & 1.82015 \\
                        0.7 & 2.20903 \\
                        0.8 & 2.88417 \\
                        0.9 & 4.47667 \\
                    \end{tabular}
                \end{tabular}
            \end{center}
            \captionof{table}{The Age of the Universe (in the unit of $(H^0)^{-1}$) for different values of $\beta$.}
        \end{minipage}
        
    \end{figure}

\noindent 
    The Age of the Universe (AOU) is the difference between the times when the scale factor is 1 (i.e., the present day) and when the scale factor was 0 (the Big Bang). This time difference which gives the AOU can be directly calculated as the definite integral of the equation 
    
    \begin{equation*}
        t_{AOU} = (H^0)^{-1} H^0\sqrt{\frac{3c^2}{8\pi G\rho_m^0}}\bigints_0^1 \frac{dx}{  x \sqrt{\displaystyle \frac{1}{x^{\gamma}} \left(1 + \frac{3\beta}{\gamma} \right) +\left(\frac{\rho_\phi^0}{\rho_m^0} -\frac{3\beta }{\gamma}\right) }  ~}
    \end{equation*}
    Using the values of Hubble's constant $H^0$, and the matter and dark energy density from cosmological observations, and integrating the equation over $x$, we obtain the age of the universe as a function of the coupling constant $\beta$
    \begin{equation}
        t_{AOU}(\beta)= (H^0)^{-1}\frac{1.21 \, _2F_1(0.5,0.5;1.5;3.33 \beta-2.33)}{\sqrt{1 -\beta}},
    \end{equation}
    for $\beta < 1$. This equation is plotted to visualize the variation of the age of the universe with the coupling constant $\beta$ in Fig.(\ref{fig:AOUvsbeta}). \\

    For $\beta \ge 1$ the definite integral does not converge to a single value and hence the age of the universe becomes indeterminate. Therefore such a model cannot represent any real universe, and hence we obtain an upper limit on the coupling constant $\beta$ for a real universe, i.e. $\beta<1$.
    \vspace{5pt}
    There does not exist a lower limit on the coupling constant, as there are no breaks or discontinuities in the age of the universe as a function of $\beta$. The $t_{AOU}$ is a smooth and continuous function of $\beta$, with a lower limit $-\infty$ and an upper limit $1$ on the parameter $\beta$. 
   The age of the universe converges to 0 as the coupling constant goes to $-\infty$, i.e. for a universe with coupling constant $\beta \rightarrow -\infty$, the evolution of the universe from $a = 0$ to $a = 1$ takes place instantaneously.
    As $\beta \rightarrow 1$, the AOU tends to $\infty$, which implies that the dynamics of such a universe are very slow on cosmic scales. This behavior is also visible in the evolution of scale factor, where the scale factor tends to 0 even at large t as $\beta \rightarrow 1$.
    The numerical value of the Age of the Universe in the terms of inverse Hubble's constant for different values of $\beta$ have also been tabulated in Table 1.


\section{Comparison of coupling strength: $3\beta \rho_m$ vs $\alpha \rho_\phi$ }\label{sec: aplha vs beta}
\noindent
The interaction term $Q$ can have two possible forms: $Q=\alpha \rho_\phi$, or $Q=3\beta \rho_m$.
The complete analysis for the former form has been done in \cite{kundu2021interacting} with a significant conclusion that even though the coupling strength has no lower bound, the upper bound on the coupling constant $\alpha$ must be 1. In this article, we did the entire analysis by taking the later form of the interaction. We obtain a similar conclusion for the coupling strength $\beta$, i.e., it has no lower bound, but the upper bound must be 1.

\begin{figure}[h]
    \centering
    \includegraphics{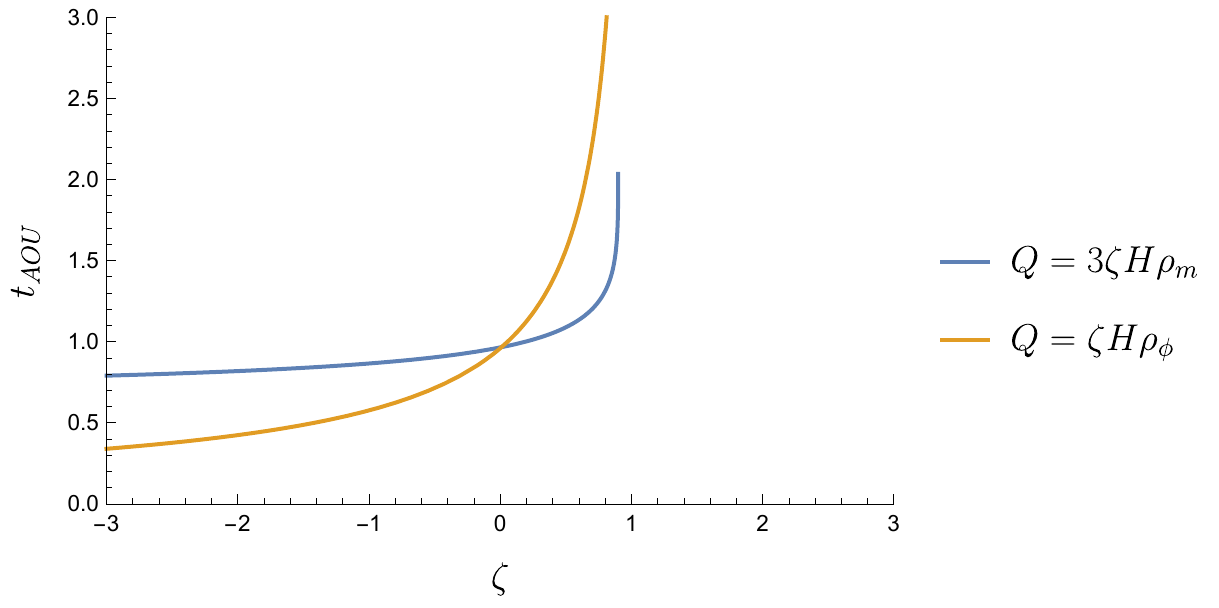}
    \caption{The age of the universe as a function of the coupling constant in the two different models considered. Here $\zeta$ is used as a common coupling constant in place of $\alpha$ and $\beta$.}
    \label{fig:comparision}
\end{figure}

As the possible range of both $\alpha$ and $\beta$ are the same $(-\infty,1)$, the question arises: are they different, or are they just the same thing with different notations? Further investigations in this line can be interesting and significant if we can replace the multiple coupling constant with just one. The age of the universe as a function of the coupling constant obtained
from both form of interaction is compared in Fig.(\ref{fig:comparision}). 

Both the coupling constants give rise to the universe, whose age decreases with the value of the coupling constant and goes to infinity as it approaches 1. Both are continuously increasing functions, and for the interaction-less universe, both converge to the same value: our universe's age.

    \section{Conclusion}\label{sec: Con}
    \noindent
   We have investigated the dynamics of a universe with an interacting tachyonic scalar field as a possible source of dark energy. The interaction between the matter and dark energy component is modeled as an energy exchange, with the energy exchange depending linearly on the energy density of the matter.
    We also presented the evolution of energy densities with respect to scale factor, as well as the evolution of density parameter ($\Omega$) with the natural logarithm of the scale factor and derived the Age of the Universe was derived as the function of the coupling constant $\beta$.

    Our analysis reveals that the case where $\beta = 0$, (no interaction between matter and TSF) corresponds to the dynamics of our universe, as expected. We have also found that for $\beta \geq 1$, the age of the universe does not converge and hence becomes indeterminate. This puts an upper bound on the possible values of the coupling constant $\beta$ for real universes. This result is similar to the case when interaction term depend linearly on the energy density of the TSF \cite{kundu2021interacting}. 
    %
%
    A comparison with the work \cite{kundu2021interacting} shows that in both the cases, the constraints on the coupling constant are the same, implying that the bound on the coupling constant is the same for both the case, irrespective of whether the interaction term depends on the energy density of matter or the energy density of dark energy (tachyonic field). Since the bound on both $\alpha$ and $\beta$ are the same, i.e. $(-\infty,1)$, it would be interesting to further investigate whether the two can be unified and replaced by a single coupling constant.
    

    

\bibliography{CosmologyLiterature.bib}
\bibliographystyle{hieeetr}
\end{document}